\documentclass[twocolumn]{aastex62}
\usepackage{graphicx}
\usepackage{color}
\usepackage{amsmath}
\usepackage{lineno}
\usepackage[first=0,last=9]{lcg}

\usepackage{array,multirow}

\definecolor{LightCyan}{rgb}{0.88,1,1}

\usepackage{appendix}
\usepackage{tabularx}
\usepackage{longtable}
\usepackage{enumerate}
\usepackage{enumitem}
\begin{document}
%GW170817+ 
%GW190521 (e=0.67) limits : %$67.685^4.569_-6.854$
%GW190521 (e=0) limits : %$43.317^23.988_-23.226$
%GW190521 (e=0.67) limits : %$89.008^19.419_-32.745$
%GW170817 limits : $73.778^8.758_-11.042$
%\preprint{APS/123-QED}

\title{Hubble Constant Measurement with GW190521 as an Eccentric Black Hole Merger}% Force line breaks with \\

\author{V. Gayathri}
\affiliation{Department of Physics, University of Florida, PO Box 118440, Gainesville, FL 32611-8440, USA}
\author{J. Healy}
\affiliation{Center for Computational Relativity and Gravitation, Rochester Institute of Technology, Rochester, NY 14623, USA}
\author{J. Lange}
\affiliation{Center for Computational Relativity and Gravitation, Rochester Institute of Technology, Rochester, NY 14623, USA}
\author{B. O'Brien}
\affiliation{Department of Physics, University of Florida, PO Box 118440, Gainesville, FL 32611-8440, USA}
\author{M. Szczepanczyk}
\affiliation{Department of Physics, University of Florida, PO Box 118440, Gainesville, FL 32611-8440, USA}
\author{I. Bartos}
\thanks{imrebartos@ufl.edu}
\affiliation{Department of Physics, University of Florida, PO Box 118440, Gainesville, FL 32611-8440, USA}
\author{M. Campanelli}
\affiliation{Center for Computational Relativity and Gravitation, Rochester Institute of Technology, Rochester, NY 14623, USA}
\author{S. Klimenko}
\affiliation{Department of Physics, University of Florida, PO Box 118440, Gainesville, FL 32611-8440, USA}
\author{C. Lousto}
\affiliation{Center for Computational Relativity and Gravitation, Rochester Institute of Technology, Rochester, NY 14623, USA}
\author{R. O'Shaughnessy}
\thanks{rossma@rit.edu}
\affiliation{Center for Computational Relativity and Gravitation, Rochester Institute of Technology, Rochester, NY 14623, USA}

\begin{abstract}
Gravitational wave observations can be used to accurately measure the Hubble constant $H_0$ and could help understand the present discrepancy between constraints from Type Ia supernovae and the cosmic microwave background. Neutron star mergers are primarily used for this purpose as their electromagnetic emission can be used to greatly reduce measurement uncertainties. Here we estimate $H_0$ using the recently observed black hole merger GW190521 and its candidate electromagnetic counterpart found by ZTF using a highly eccentric explanation of the properties of GW190521. We find that the reconstructed distance of GW190521 and the redshift of the candidate host galaxy are more consistent with standard cosmology for our eccentric model than if we reconstruct the source parameters assuming no eccentricity. We obtain $H_0=88.6^{+17.1}_{-34.3}$\,km\,s$^{-1}$Mpc$^{-1}$ for GW190521, and $H_0=73.4^{+6.9}_{-10.7}$\,km\,s$^{-1}$Mpc$^{-1}$ in combination with the results of the neutron star merger GW170817. Our results indicate that future $H_0$ computations using black hole mergers will need to account for possible eccentricity. For extreme cases, the orbital velocity of binaries in AGN disks can represent a significant systematic uncertainty.
\end{abstract}

%\maketitle

%\tableofcontents

%%%%%%%%%%%%%%%%%
\section{Introduction}
%%%%%%%%%%%%%%%%%

With a total mass of around $150\,$M$_{\odot}$, the binary black hole merger GW190521 was the heaviest system detected to date through gravitational waves by LIGO and Virgo \citep{2015CQGra..32g4001L,2015CQGra..32b4001A,GW190521discovery}. The heavier black hole in the binary had a mass of about $85$\,M$_\odot$. Such a mass is not expected from stellar evolution due to pair instability that prevents some of the most massive stars from leaving a compact remnant \citep{Woosley2007,GW190521properties}. In addition, the black holes' spins are found to be large and misaligned with the binary orbit, disfavoring the possibility that the system is originated from a stellar binary \citep{GW190521properties}.

A possible explanation for the observed properties of GW190521 is that it is a so-called hierarchical merger---the black holes in the binary are themselves the remnants of past black hole mergers \citep{MillerHierarchical2002,O_Leary_2006,Giersz2016}. This scenario can naturally lead to masses in excess to the $\sim65$\,M$_\odot$ pair-instability limit. It also results in higher black hole spins. The merger of two black holes with the same mass and no spin will produce a remnant black hole with 0.7 dimensionless spin \citep{Lousto:2009ka}, consistent with the reconstructed spins of $0.69^{+0.27}_{-0.62}$ and $0.73^{+0.24}_{-0.64}$ for the two black holes in GW190521 \citep{GW190521discovery}. In addition, the hierarchical merger scenario implies that black holes form a binary after a chance encounter, in which their spin will be randomly oriented. this is consistent with the reconstructed misalignment between the binary orbit and black hole spins in GW190521 \citep{GW190521discovery}.

By comparing the observed gravitational waveform to numerical relativity simulations, \cite{GW190521_eBBH} found that GW190521 is probably a highly eccentric merger (hereafter UF/RIT model). This result further supports the binary's origin as a dynamical encounter within a dense black hole population. Binaries lose any existing eccentricity over time due to gravitational radiation, therefore only binaries that formed soon before merger can retain any eccentricity. Such formation is possible in chance encounters but not in systems originating in isolated stellar binaries.

Active Galactic Nuclei (AGNs) represent a well-suited environment to produce hierarchical black hole mergers \citep{2017ApJ...835..165B,2017NatCo...8..831B,2017MNRAS.464..946S,2018ApJ...866...66M,2019PhRvL.123r1101Y,2019ApJ...876..122Y,2020ApJ...898...25T,2020ApJ...899...26T,Yang_2020}. Galactic nuclei harbor a dense population of black holes \citep{2009MNRAS.395.2127O,2018Natur.556...70H} which are further compressed through interaction with the AGN disk. Dynamical friction will align the orbits of some of the black holes with the disk plane, where they migrate inward and can merge with each other. As merger remnants can remain within the disk, consecutive mergers are common and could represent the majority of AGN-assisted mergers. Several black hole mergers previously discovered by LIGO/Virgo have properties suggestive of their possible AGN origin \citep{2019PhRvL.123r1101Y,2020ApJ...890L..20G,2020arXiv200704781Y}. 

Following the public alert issued by LIGO/Virgo on the detection of GW190521 \citep{GW190521_GCN}, the Zwicky Transient Facility (ZTF) carried out a search for excess optical emission from an AGN within the publicly available localization volume of GW190521. It identified a possible counterpart that was interpreted as being due to the accreting black hole remnant of the GW190521 merger \citep{2020PhRvL.124y1102G}. As this is the first such observation and since there are open questions about the emission processes involved, more studies and probably further similar detections are needed to confidently establish the connection between the transient and GW190521. However, for the purposes of understanding the consequences of such a connection, in the following we assume that the electromagnetic emission is indeed produced by the merger remnant.

In this paper we constrained $H_0$ using GW190521 and its candidate ZTF counterpart. We used the reconstructed properties of GW190521 by the UF/RIT model in which the event was a highly eccentric black hole binary with eccentricity $e\approx0.7$ \citep{GW190521_eBBH}.

The paper is organized as follows. We describe our computation of $H_0$ in Section \ref{sec:H0}. We present our localization results and the computed $H_0$ for GW190521 in Sections \ref{sec:localiation} and \ref{sec:results}, respectively. We conclude in \ref{sec:conclusion}.

%%%%%%%%%%%%%%%%%
\section{Source localization}
\label{sec:localiation}
%%%%%%%%%%%%%%%%%

\begin{figure}
\vspace{1cm}
  \includegraphics[angle=0,width=1\columnwidth]{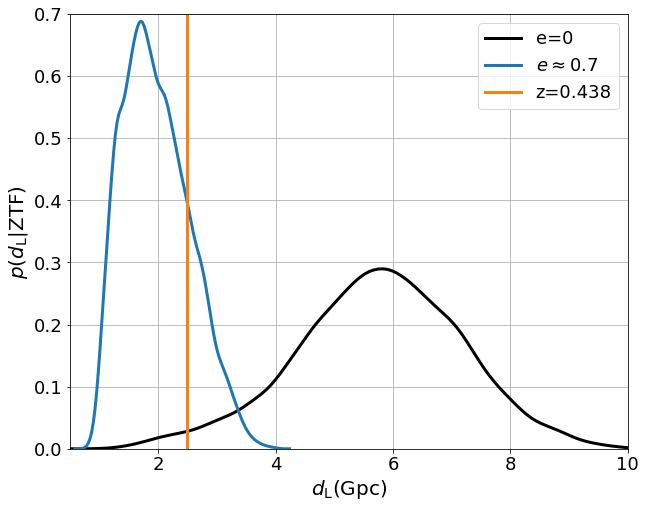}
      \caption{{\bf Luminosity distance probability distribution} obtained using NRSur7dq4 gravitational waveforms \citep{PhysRevResearch.1.033015} assuming eccentricity $e=0$ (red), and using the UF/RIT model with eccentricity $e\approx0.7$ \citep{GW190521_eBBH} (black). These distributions are obtained using RIFT algorithm for fixed source direction to that of the ZTF source. The vertical line shows the distance of the ZTF source assuming Planck 2018 cosmology \citep{collaboration2018planck}.
    \label{fig:Distance}}
\end{figure}

We used the recorded gravitational wave data $\mathcal{D}_{\rm GW}$ to compute the probability density $p(d_{\rm L}|\mathcal{D}_{\rm GW})$ of the merger's luminosity distance $d_{\rm L}$. We used the RIFT parameter estimation package \citep{gwastro-PENR-RIFT,gwastro-PENR-Methods-Lange,NRPaper,2018PhRvD..97f4027H} to obtain this probability density. We fixed the source direction to that of the ZTF candidate, $\Omega_{\rm ZTF}$. We carried out this computation for both our UF/RIT model with $e\approx0.7$ obtained using numerical relativity simulations \citep{GW190521_eBBH}, and for a non-eccentric model derived using NRSur7dq4 waveforms \citep{PhysRevResearch.1.033015}. For both models we assumed that the electromagnetic counterpart is always detectable from this source type independently from the source direction and distance. We adopted a uniform volumetric source probability density, which is a good approximation of the expected distribution of AGN-assisted mergers \citep{Yang_2020}. We further adopted a uniform prior on the cosine of the binary's inclination. For our $e\approx0.7$ model we adopted mass and spin parameters from the maximum-likelihood waveform  \citep{GW190521_eBBH}. For our $e=0$ model we used uniform probability densities for the black hole masses within $[30M_\odot,200M_\odot]$, uniform spin amplitudes and isotropic spin orientations. Finally, we neglected selection effects related to the $H_0$-dependence of the source population density at $z_{\rm ZTF}$, which are not expected to be significant at $z\sim0.4$ \citep{Farrgit}.

In Fig. \ref{fig:Distance} we show $p(d_{\rm L}|\mathcal{D}_{\rm GW},\Omega_{\rm ZTF})$ for the UF/RIT model with $e\approx0.7$ \citep{GW190521_eBBH} and also one derived using the NRSur7dq4 waveform model \citep{PhysRevResearch.1.033015} assuming $e=0$. We find that the distributions are markedly different for these two cases (see also \citealt{2020arXiv200901066C}). %While \cite{GW190521discovery} reconstructs a distance of $d_{\rm}(\mbox{NRSur7dq4})=5.3^{+1.5}_{-1.5}$\,Gpc, \cite{GW190521_eBBH} obtains $d_{\rm}(e\approx0.7)=1.8^{+1.1}_{-0.1}$\,Gpc. We see that the difference is significant and not compatible within the error bars.

%With this extra information we find a distance of $d_{\rm}(e\approx0.7,\mbox{fixed sky})=1.9^{+0.4}_{-0.7}$\,Gpc and for $d_{\rm}(NRSur7dq4,\mbox{fixed sky})=4.7^{+1.2}_{-2.9}$. See the probability density for this "fixed-sky" model in Fig. \ref{fig:Distance}. 
%\sout{We used this fixed-sky distance distribution for our $H_0$ estimate below} \Carlos{Do we need this sentence?}.

\begin{figure}
\vspace{0.5cm}
  \includegraphics[angle=0,width=1\columnwidth]{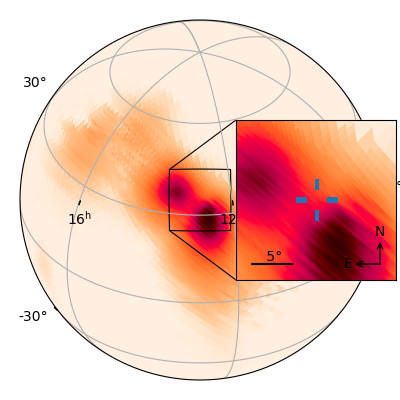}
      \caption{{\bf Sky location reconstructed for GW190521 using the UF/RIT model with $e\approx 0.7$ \citep{GW190521_eBBH}.} In the inset panel, the reticle marks the position of the apparent host AGN candidate J124942.3+344929 \citep{2020PhRvL.124y1102G}.
    \label{fig:location}}
\end{figure}

The ZTF candidate counterpart was associated with AGN J124942.3+344929 with measured redshift $z_{\rm ZTF} = 0.438$ \citep{2020PhRvL.124y1102G}. In Fig. \ref{fig:location} we also show the distance of the ZTF candidate assuming Planck 2018 cosmology \citep{collaboration2018planck}. We see that both $e\approx0.7$ and $e=0$ models are consistent with this distance, with somewhat higher probability density for the eccentric case.

\newpage

%%%%%%%%%%%%%%%%%
\section{Computing the Hubble Constant}
\label{sec:H0}
%%%%%%%%%%%%%%%%%

The Hubble constant $H_0$ describes the local expansion rate of the Universe. This explains the motion of astronomical objects due to this expansion. It is expressed as $v_{\rm h}=H_0 d_{\rm L}$, where $v_{\rm h}$ is Hubble flow velocity and $d_{\rm L}$ is the luminosity distance to the source. 

Gravitational waves from compact binary mergers enable us to directly measure the luminosity distance of the source. If we are able to identify a binary's host galaxy, the host galaxy provides information on the binary's redshift. As gravitational wave localization is typically limited, the host galaxy identification relies primarily on the detection of electromagnetic emission from the binary. Once we have an estimated redshift for the source, for fixed $d_{\rm L}$ one can estimate the Hubble constant \citep{Hogg:1999ad} using
%\begin{equation}
%    H_0= \frac{2c\left(2-\Omega_m (1-z)-(2-\Omega_m)\sqrt{1+\Omega_m z}\right)}{d_{\rm L}\Omega_m^2}
%\end{equation}
\begin{equation}
   H_0(d_{\rm L},z)=\frac{c(1+z)}{d_{\rm L}}\int_0^{z}\frac{dz'}{E(z')}
   \label{eq:dist}
\end{equation}
with
\begin{equation}
E(z)=\sqrt{\Omega_{\rm r}(1+z)^4+\Omega_{\rm m}(1+z)^3+\Omega_{\rm k}(1+z)^2+\Omega_{\rm \Lambda}}\,.
\end{equation}
Here, $c$ is the speed of light, $\Omega_{\rm r}$ is the radiation energy density, $\Omega_{\rm m}$ is matter density, $\Omega_{\rm \Lambda}$ is the dark energy density and $\Omega_{\rm k}$ is the curvature of our Universe. We adopted a set of cosmology parameters $\{H_0, \Omega_{\rm r}, \Omega_{\rm m}, \Omega_{\rm k}, \Omega_{\rm \Lambda}\}=\{67.8{\rm km\ s^{-1}\ Mpc^{-1}},\ 0,\ 0.306,\ 0,\ 0.694\}$ measured by the Planck satellite \citep{collaboration2018planck}. We considered these parameters fixed when recovering $H_0$ as their uncertainties are much smaller than other uncertainties here. Given the Gpc distance scale of the event, we neglected the peculiar velocity of the host galaxy. 

We also neglected any motion of the binary within the galaxy. Considering the mass of the supermassive black hole in the candidate AGN, $M_{\rm SMBH}=10^8-10^9$\,M$_\odot$ \citep{2020PhRvL.124y1102G}, and the characteristic distance $10^{-2}$\,pc of the merger from the supermassive black hole, the rotational velocity of the binary is $10^4$\,km\,s$^{-1}$. For the reconstructed distance of GW190521, this is a $1-30\,\%$ error on the reconstructed Hubble constant depending on the orientation of the AGN disk plane and the mass and luminosity distance of the supermassive black hole. This is smaller than the statistical error here, but will need to be examined more carefully if a larger number of AGN-assisted binaries are used to measure $H_0$. 

We computed the probability density of the Hubble constant  using the distance probability density:
%\begin{widetext}
\begin{eqnarray}\nonumber
p(H_0| \mathcal{D}_{\rm GW}, \Omega_{\rm ZTF}, z_{\rm ZTF}) = \\ p\left(d_{\rm L}(H_0,z_{\rm ZTF})| \mathcal{D}_{\rm GW}, \Omega_{\rm ZTF}\right) \frac{\partial}{\partial d_{\rm L}}H_0(d_{\rm L},z_{\rm ZTF})
\end{eqnarray}
%\end{widetext}
where $d_{\rm L}(H_0,z_{\rm ZTF})$ is the inverse function of Eq. \ref{eq:dist}.

For comparison, see \cite{MukherjeeH0} and \citep{ChenH0} who also computed $p(H_0| \mathcal{D}_{\rm GW}, \Omega_{\rm ZTF}, z_{\rm ZTF})$ for GW190521 for $e=0$.

%%    term1=2.0
  % term2=omega_om*(1.0-z)
   % term3=(2.0-omega_om)*np.sqrt(1+omega_om*z)
    %term4=omega_om**2#*(1+z)**2
    %c=299792.46
    %print(2*((term1-term2-term3)*term4**(-1)))
    %H0=(c/Dl)*2*((term1-term2-term3)*term4**(-1))
%With this procedure we have already constrained the Hubble constant using the binary neutron star merger GW170817 \citep{hubble_gw170817}.  

%In $21^{st}$ May 2019, LIGO/Virgo detected GW signal from intermediate mass black hole binary and named as GW190521 \cite{detection_gw190521.1} with source total mass $X$ and luminosity distance $Y Gpc$. In our recent work we have estimated source parameters for this event if it is eccentric binary black hole. We have reported the estimated total mass as $XX$, luminosity distance $ 2.49 ^{+ 3.48 }_{ -1.38 } Gpc$ and eccentricity $e=0.67$. Recently, EM group reported EM contour part associated with this GW event at red-shift $z=0.438$ \cite{Graham:2020gwr} and sky-location (right ascension, declination) is ($192.26 , 34.82$). The estimated red-shift corresponds to distance $~2.4 Gpc$ .   

%%%%%%%%%%%%%%%%%
\section{Results}
\label{sec:results}
%%%%%%%%%%%%%%%%%

\begin{figure}
\vspace{0.5cm}
  \includegraphics[angle=0,width=1\columnwidth]{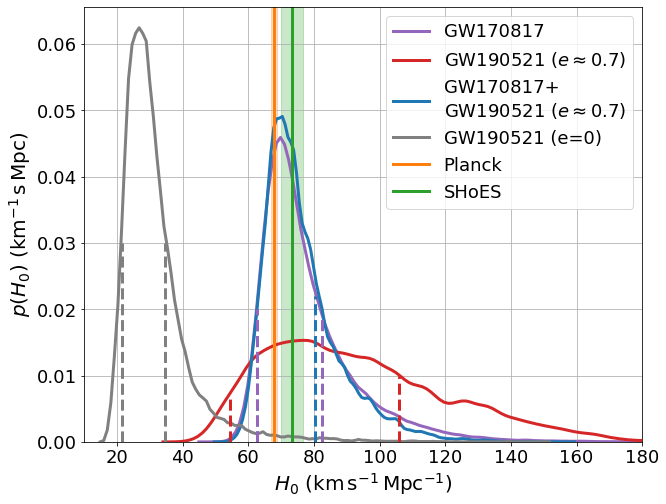}
      \caption{{\bf $H_0$ measurements for GW190521} with its ZTF candidate counterpart and GW170817. The following $H_0$ probability densities are shown: GW170817 (purple); GW190521 with eccentric model (red); combined GW170817 and GW190521 with eccentric model (blue); GW190521 with $e=0$ (gray); cosmic microwave background results by Planck (orange); and type Ia supernova results by ShoES (green). Shaded areas for the latter two results show 95\% confidence intervals. Vertical dashed lines for the gravitational-wave results indicate 68\% credible intervals.
    \label{fig:h0}}
\end{figure}

Our $H_0$ probability density from GW190521 based on the UF/RIT model \citep{GW190521_eBBH} and the ZTF candidate counterpart is shown in Fig. \ref{fig:h0}. Numerically it is $H_0=88.6^{+17.1}_{-34.3}$\,km\,s$^{-1}$Mpc$^{-1}$. For comparison we show our $H_0$ estimate for $e=0$. We see that the distribution from the eccentric model has its maximum near the $H_0$ values measured using type Ia supernovae, which give a local expansion rate of $H_0 = 74.03 \pm 1.42$\,km\,s$^{-1}$\,Mpc$^{-1}$  \citep{2019ApJ...876...85R}, and the estimate from cosmic microwave background observations measured by the Planck satellite, which gives $H_0 = 67.4\pm 0.5$\,km\,s$^{-1}$\,Mpc$^{-1}$ \citep{collaboration2018planck}. The uncertainty is nevertheless significant. Our $e=0$ result is also consistent with the Type Ia / Planck $H_0$ estimate.

We also show in Fig. \ref{fig:h0} the expected $H_0$ estimate after combining the distributions obtained for GW190521 with that of GW170817. We see that the improvement by this combination, as measured by the height of the probability density distribution, is a few percent, i.e most information still comes from GW170817.

%%%%%%%%%%%%%%%%%%%%%%%%%%%
\section{Conclusion}
\label{sec:conclusion}
%%%%%%%%%%%%%%%%%%%%%%%%%%%

We estimated the Hubble constant using the luminosity distance of the gravitational wave signal GW190521 and the redshift of its candidate electromagnetic counterpart detected by ZTF, assuming that the association is real. For GW190521 we used the highly eccentric UF/RIT model \citep{GW190521_eBBH} with $e\approx0.7$, and for comparison a non-eccentric model similar to that of \cite{GW190521discovery}. Our conclusions are as follows.
\begin{itemize}[leftmargin=*,itemsep=2pt,parsep=2pt]
\item We find $H_0=88.6^{+17.1}_{-34.3}$\,km\,s$^{-1}$Mpc$^{-1}$ for GW190521 as a highly eccentric merger with $e\approx0.7$. 
%\item Eccentricity increases the consistency of Assuming no eccentricity \citep{GW190521discovery}, the obtained $H_0\sim36$\,km\,s$^{-1}$Mpc$^{-1}$ is also consistent with  only marginally consistent with standard results.
\item Combining GW190521 and GW170817, we find $H_0=73.4^{+6.9}_{-10.7}$\,km\,s$^{-1}$Mpc$^{-1}$. 
\item $H_0$ measurements using black hole mergers could be strongly affected if eccentricity is present and is not accounted for.
\item $H_0$ measurements using multiple AGN-assisted black hole mergers or mergers in galactic nuclei need to consider the effect of Doppler shift due to the binary's orbital velocity that in extreme cases can introduce a large systematic uncertainty.
\end{itemize}

The authors are grateful to Will Farr for useful suggestions. The authors gratefully acknowledge the National Science Foundation (NSF)
for financial support from Grants
No.\ PHY-1912632, No.\ PHY-1806165, No.\ PHY-1707946, No.\ ACI-1550436, No.\ AST-1516150,
No.\ ACI-1516125, No.\ PHY-1726215, NSF AST-1516150, PHY-1707946, NASA TCAN grant No. 80NSSC18K1488, and the support of the Alfred P. Sloan Foundation.
This work used the Extreme Science and Engineering Discovery Environment (XSEDE) [allocation TG-PHY060027N], which is supported by NSF grant No. ACI-1548562 and Frontera projects PHY-20010 and PHY-20007.
Computational resources were also provided by the NewHorizons, BlueSky Clusters, and Green Prairies at the Rochester Institute of Technology, which were supported by NSF grants No.\ PHY-0722703, No.\ DMS-0820923, No.\ AST-1028087, No.\ PHY-1229173, and No.\ PHY-1726215. We are grateful for computational resources provided by the Leonard E Parker Center for Gravitation, Cosmology and Astrophysics at the University of Wisconsin-Milwaukee supported by NSF Grant No. PHY-1626190.  
This research has made
use of data, software and/or web tools obtained from
the Gravitational Wave Open Science Center (https:
//www.gw-openscience.org), a service of LIGO Laboratory, the LIGO Scientific Collaboration and the Virgo
Collaboration. 
LIGO is funded by the U.S. National Science Foundation. Virgo is funded by the French Centre National de Recherche Scientifique (CNRS), the Italian
Istituto Nazionale della Fisica Nucleare (INFN) and the Dutch Nikhef, with contributions by Polish and Hungarian institutes.

\bibliography{references}
\end{document}